\documentclass[11pt]{article} 
\usepackage{graphics} 
\usepackage{epsfig} 
\usepackage{amsmath} 
\usepackage{color} 
\usepackage{caption} 
\usepackage{makecell} 
\usepackage{multirow} 
\usepackage{longtable} 
\usepackage{rotating} 
\usepackage[leftcaption]{sidecap} 
\usepackage[TABBOTCAP]{subfigure} 
\usepackage{epstopdf} 
\usepackage{lineno} 
\usepackage{booktabs} 
\usepackage{authblk} 
\newcommand{\mage} {\textsc{MaGe}}

\begin{document} 
\renewcommand{\arraystretch}{1.2}

\title{Signal recognition efficiencies of artificial neural-network pulse-shape discrimination in HPGe $0\nu\beta\beta$-decay searches} 
\author{A.~Caldwell, F.~Cossavella, B.~Majorovits, D.~Palioselitis, O.~Volynets} \affil{Max-Planck-Institut f{\"u}r Physik, F{\"o}hringer Ring 6, 80805 M{\"u}nchen, Germany}

\maketitle
\begin{abstract}
	A pulse-shape discrimination method based on artificial neural networks was applied to pulses simulated for different background, signal and signal-like interactions inside a germanium detector. The simulated pulses were used to investigate variations of efficiencies as a function of used training set. It is verified that neural networks are well-suited to identify background pulses in true-coaxial high-purity germanium detectors. The systematic uncertainty on the signal recognition efficiency derived using signal-like evaluation samples from calibration measurements is estimated to be 5\%. This uncertainty is due to differences between signal and calibration samples. 
\end{abstract}

\section{Introduction}\label{intro}Experiments searching for neutrinoless double beta ($0\nu\beta\beta$) decay require an extremely low background level in the region of interest around a few MeV. Compton scattered $\gamma$-particles, originating from radioactive decays in the proximity of the detectors, are an important background contribution at such energies. 

In high-purity germanium (HPGe) experiments, these interactions are often identified and removed from the signal data set through pulse-shape analysis (PSA). In order to extract a half-life limit, the signal recognition efficiency has to be known. Usually, experimentally obtained pulse-shape libraries with signal-like events are used to obtain the signal recognition efficiency. However, these evaluation libraries can have energy-deposition topologies and event-location distributions different to those of the signal searched for. Efficiencies obtained like this can be systematically different from the recognition efficiency for the real signal. Furthermore, the evaluation libraries used to derive the recognition efficiencies often contain events of the wrong type, making a direct determination of the efficiencies impossible.

This paper presents investigations of the reproducibility and systematic uncertainties of the efficiencies of pulse-shape discrimination (PSD) using artificial neural networks (ANNs) with libraries of simulated pulses. The general idea of PSD using ANNs is introduced and the sources of possible systematic effects are discussed. The simulations and the libraries used for the analysis are described as well as the ANNs and the procedures used to train them. The stability of the method against initial conditions and ANN topologies is investigated. The focus is on the differences obtained in recognition efficiencies using different evaluation libraries and the associated systematic uncertainties.

\section{Pulse-shape discrimination for HPGe detectors using artificial neural-networks}\label{sec:1} The detection principle of semiconductor detectors is based on the creation and detection of electron--hole pairs, i.e.\ charge carriers, when radiation interacts with the detector material. Charge-sensitive preamplifiers are commonly used to detect the drifting charge carriers in large volume HPGe detectors. The time structure of an event, the pulse-shape, is defined by the mirror charge signal induced on the electrodes as a function of time. The pulse length is given by the time needed to fully collect the charges on the electrodes. See e.g.~\cite{knoll,PSS} for a detailed description of the pulse creation process. 

For photons in the MeV range, the dominant interaction process is Compton scattering. A photon with an energy of one MeV has a mean free path of $\simeq$3\,cm in germanium. Thus, photon-induced events with energies of about 2\,MeV are mostly composed of several energy deposits within a HPGe detector, separated by a few centimeters. These background-like events are referred to as multi-site events (MSE). In contrast, electrons with the same energy have a range of the order of millimeters and deposit their kinetic energy ``locally''. Signal-like events of this kind are referred to as single-site events (SSE). Note that in reality there also exists ``signal-like'' background, i.e.\ background events that have an indistinguishable event topology, such as the irreducible background from $2\nu\beta\beta$ decay. The two electrons emitted in 0$\nu\beta\beta$ decay result predominantly in SSEs. Due to Bremsstrahlung, a fraction of a few \% of the 0$\nu\beta\beta$-decay events become MSEs~\cite{Kevin_thesis}. Events identified as MSE in the energy region of interest are rejected as background. 

Methods to distinguish between SSEs and MSEs in HPGe detectors using ANNs were developed previously~\cite{Majo,kevin,gerda_psa}. In most previous works, events from double escape peaks (DEP) and full energy peaks (FEP) were used to create training libraries of signal-like and background-like events, respectively. These were obtained from calibration data for which sources such as $^{228}$Th or $^{56}$Co were used. The ANN efficiencies to correctly identify events are also typically extracted using evaluation libraries from calibration measurements. 

The efficiencies of PSD methods are not necessarily homogeneous throughout the detector volume. For a realistic evaluation, the spatial distribution of the events in a given test library has to be taken into account. Especially, DEP events will exhibit a non-uniformity in event location distribution due to the topology of the events. If pair production occurs in a coaxial HPGe at high radii, $r$, and height, $z$, i.e.\ close to the extreme boundaries, the probability for the two 511\,keV $\gamma$-particles to escape is the highest. Hence, libraries of DEP events have a higher event location density in these parts of the detector (see Section~\ref{sec:libraries} and Fig.~\ref{fig:psa:event_distr}). On the other hand, signal events due to $0\nu\beta\beta$ decay (but also ``signal-like'' background events due to $2\nu\beta\beta$ decay) are expected to be homogeneously distributed. Using a library with an event location distribution different from the one expected for the signal can lead to systematic biases. The main scope of this work is to address this issue and estimate the uncertainties on the SSE recognition efficiency evaluation arising from the use of different training and evaluation sets.

\section{Strategy} In order to quantify the uncertainties on the ANN event topology recognition efficiencies, simulations are used. The signal (background) recognition efficiency $\eta$ ($\rho$) of any PSD method is defined as the probability that the method correctly identifies an SSE (MSE) from an event-library containing only SSEs (MSEs).

Realistic SSE and MSE pulse-shape libraries always contain events of both classes. Hence, the ANN method applied to a library of predominantly SSE or MSE pulses will result in a survival probability $E$ or rejection probability $R$, defined as the fraction of pulses in the library that are classified as SSE or MSE, respectively: 
\begin{eqnarray}
	E = \eta \cdot S_{SSE} + (1-\rho) \cdot S_{MSE},\nonumber \\
	R = \rho \cdot B_{MSE} + (1-\eta) \cdot B_{SSE}, \label{equ:1} 
\end{eqnarray}
where $S_{SSE}$ and $S_{MSE}$ are the fraction of SSEs and MSEs in the SSE-library, respectively, and $B_{MSE}$ and $B_{SSE}$ are the fraction of MSEs and SSEs in the MSE-library, respectively. Using simulated pulses idealized libraries with $S_{SSE}=1$ and $B_{SSE}=0$ can be created. These libraries can be used to determine $\eta$ and $\rho$ directly, as for this case $E(S_{SSE}=1, S_{MSE}=0)=\eta$ and $R(B_{MSE}=1, B_{SSE}=0)=\rho$ (see equ.~\ref{equ:1}).

In order to quantify the effect of non-homogeneous event-location distributions, $\eta$ for ANNs obtained with training libraries with inhomogeneous event-location distributions are compared to those obtained from libraries with a homogeneous event-location distribution. 

The stability of the method is verified by training and evaluating a set of ANNs with different initial weights of the ANN synapses and with training libraries of different sizes. Finally, the influence of the number of hidden layers and the number of neurons in the ANN on $\eta$, is investigated.

Signal recognition efficiencies obtained using evaluation libraries with event location distributions as expected and different from the signal are then compared. 

True-coaxial HPGe detectors are considered in this paper. They have a simple radial electric field and, thus, have relatively simple pulse shapes. Consequently, pulse-shapes of this type of detectors have lower systematic uncertainty due to smaller uncertainties in the field calculations compared to detectors with more complex geometries. This makes this type of detectors interesting for this analysis. 

\section{Libraries of simulated pulse shapes}\label{sec:libraries} HPGe detectors for low background experiments typically have a radius, $r_{\max}$, and a height of a few cm\footnote{A polar coordinate system is used with the origin at the center of the crystal and the $z$ axis pointing upwards. In Cartesian coordinates, the $x$- and $y$- axes coincide with the crystallographic $\langle110\rangle$ axes, while the $z$ axis coincides with the crystallographic $\langle001\rangle$ axis.}. The simulated n-type true-coaxial germanium detector has a height of 70\,mm and $r_{\max}$=37.5\,mm with the diameter of the borehole being 10\,mm. The dead layer due to the n+ contact (outer surface) is less than 1\,$\mu$m, while the dead layer due to the p+ contact is 0.5\,mm. The simulated geometry describes an existing true-coaxial 18-fold segmented n-type HPGe developed as a prototype detector~\cite{segmented} for the GERDA experiment~\cite{gerda}. 

Photon and electron interactions for different libraries were simulated within the \mage\ framework~\cite{MaGe}, based on Geant4~\cite{geant4,geant4_2}. Pulse shapes were simulated for the core electrode. Whenever individual energy deposits within one event were separated by less than 0.1\,mm they were combined. Pulse shapes for the combined energy deposits were simulated using pre-calculated electric and weighting fields using the pulse-shape simulation package described in~\cite{PSS}. The charge collection efficiency is either zero or one. Charge cloud diffusion and self repulsion effects are not taken into account in the simulation. The drift path anisotropy originating from the axis effect, i.e.\ the dependence of mobilities on the axis orientation, is accounted for in all simulations. 

The number of grid points for the electric- and weighting-field calculations was $33(r)\times181(\phi)\times71(z)$. The electrically active impurities were assumed to be homogeneous within the detector, with a density of $0.63\times10^{10}\,\text{cm}^{-3}$. The length of the simulated pulses is 1\,$\mu$s. The step frequency of the simulation is 1125\,MHz, a multiple of 75\,MHz to which the pulses were resampled to take the effects of a typical DAQ into account. Above 1\,GHz, the step frequency is sufficient to correctly describe trajectories~\cite{Daniel,Jing}.

\begin{figure*}
	\centering \subfigure[]{ \epsfig{file=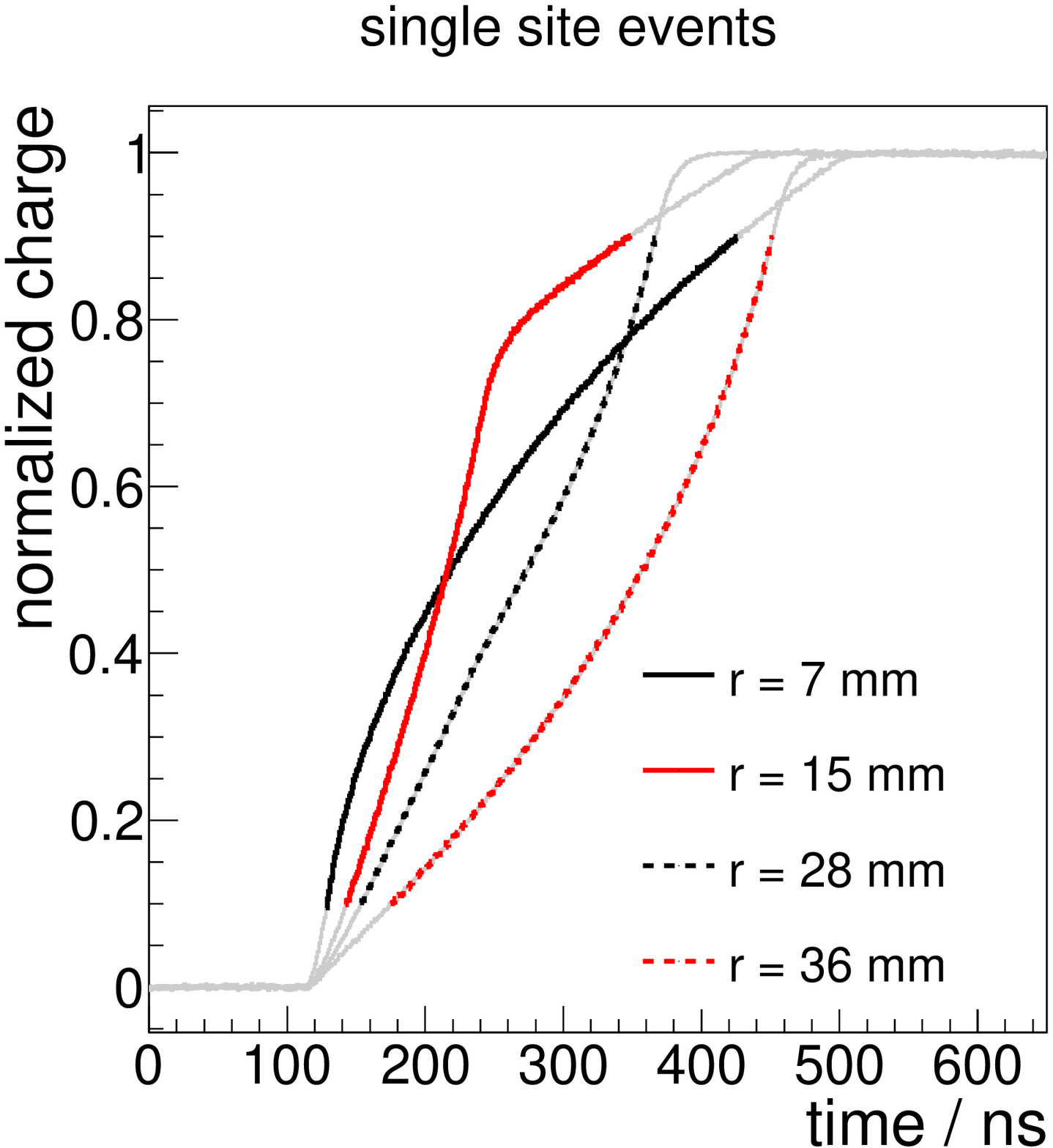,width=0.45 
	\textwidth}\label{fig:sse} } \subfigure[]{ \epsfig{file=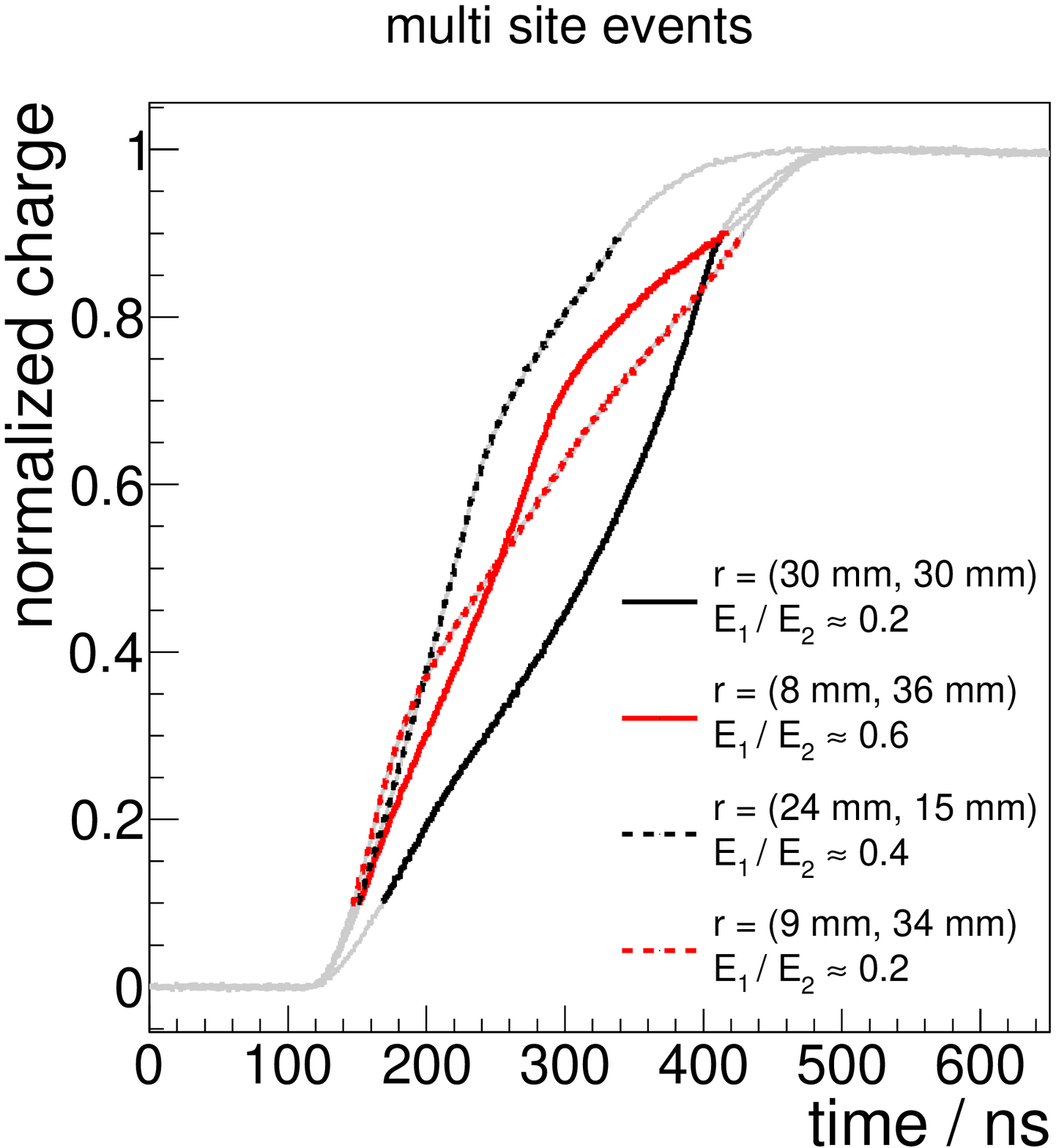,width=0.45 
	\textwidth}\label{fig:mse} }
	
	\caption{Pulse shapes for (a) SSEs and (b) MSEs\@ corresponding to $0\nu\beta\beta$ and FEP events, respectively. The colored part of the pulse indicates the time during which the amplitude is between 10\% and 90\% of the maximum amplitude. The displayed SSE pulses correspond to events at different radial positions, $r$. The MSE pulses correspond to events with two main energy depositions, $E_1$ and $E_2$, with different radial positions and energy ratios $E_1/E_2$.}\label{fig:sample_pulse} 
\end{figure*}

The amplifier RC-integration constant was set to 20\,ns, corresponding to a bandwidth of about 10\,MHz, while the amplifier decay time was set to 50\,$\mu$s. Each individual pulse shape was convoluted with Gaussian noise, $\sigma=6$\,keV. The results presented in this work do not change when simulated pulses with no noise are used, i.e.\ the efficiencies obtained are within the uncertainties quoted in the following. SSE and MSE pulses take on a wide variety of shapes as shown in Fig.~\ref{fig:sample_pulse}. It is not trivial to interpret the pulse shapes and distinguish between SSEs and MSEs without an involved quantitative analysis. The pulse length, $t_r^{10\text{--}90}$, is between 160 and 500\,ns~\cite{kevin,aleks_thesis}, where $t_r^{10\text{--}90}$ is defined as the time in which the pulse increases from 10\% to 90\% of its amplitude. This part of the pulse contains the relevant information regarding the event topology. 

Training and evaluation libraries with independent pulses were created. The simulated libraries are listed in Table~\ref{tab:libraries}.

The DEP, 2$\nu\beta\beta$ and 0$\nu\beta\beta$ event-libraries were created with and without a realistic admixture of MSEs due to Bremsstrahlung and Compton-scattered $\gamma$-particles. All MSE libraries were simulated for the 1620\,keV FEP, corresponding to a $^{228}$Th source, typically used for calibration. The notation of \textit{No Comp} \& \textit{Brems} is used for SSE libraries in which all events with Compton scattering or hard Bremsstrahlung were removed. MSE libraries that contain only events which have at least one energy deposition due to Compton scattering or hard Bremsstrahlung in the detector and thus have at least two distinct energy deposits are marked as \textit{Comp} \& \textit{Brems only}. In order to obtain a clean MSE library it was required that R$_{90}$, the radius within which 90\% of the deposited energy was contained~\cite{kevin}, is larger than 2\,mm. This ensures that all events have at least two energy deposits that are at least 2\,mm apart.

To indicate the origin of incoming photons, the last column in Table~\ref{tab:libraries} lists either ``Top'', ``Side'' or ``Homog''. This means that the photons were simulated to come from either the $xy$- or $xz$-plane for ``Top'' and ``Side'', respectively. Their origins are homogeneously distributed on these planes with their momentum perpendicular to the plane of origin. The planes are located 17.5\,cm from the center of the detector and their area is sufficiently large to cover the detector. SSE libraries with homogeneous event location distributions within the detector volume are listed as ``Homog''. For \textit{DEP clean}, 2.6\,MeV photons were forced to make pair creation with the event vertices homogeneously distributed within the detector. Each training and evaluation library contains between 7.000 and 20.000 simulated pulses. 
\begin{table*}\scriptsize
	\centering 
	\begin{tabular}
		{lcccc} Library & Energy & Processes & Source location\\
		\midrule \multicolumn{4}{c}{\bf SSE - Single Site Event Libraries}\\
		\midrule \textit{DEP top} & (1593\,$\pm$\,5)\,keV & \textit{No Comp} \& \textit{Brems}& Top\\
		\textit{DEP side} & (1593\,$\pm$\,5)\,keV & \textit{No Comp} \& \textit{Brems} & Side\\
		\textit{DEP real } & (1593\,$\pm$\,5)\,keV & All processes& Side\\
		\textit{DEP clean } & (1593\,$\pm$\,5)\,keV & \textit{No Comp} \& \textit{Brems} & Homog.\\
		\textit{0$\nu\beta\beta$ real} & (2039\,$\pm$\,5)\,keV & All processes & Homog.\\
		\textit{0$\nu\beta\beta$ clean} & (2039\,$\pm$\,5)\,keV & \textit{No Comp} \& \textit{Brems} & Homog.\\
		\textit{2$\nu\beta\beta$ real} & 450\,keV\,--\,540\,keV & All processes & Homog.\\
		\textit{2$\nu\beta\beta$ clean} & 1000\,keV\,--\,1450\,keV & \textit{No Comp} \& \textit{Brems} & Homog.\\
		\midrule \multicolumn{4}{c}{\bf MSE - Multi Site Event Libraries}\\
		\midrule \textit{FEP top} & (1620\,$\pm$\,5)\,keV & \textit{Comp} \& \textit{Brems only} & Top\\
		\textit{FEP all} & (1620\,$\pm$\,5)\,keV & All processes & Top\\
		\textit{FEP side} & (1620\,$\pm$\,5)\,keV & \textit{Comp} \& \textit{Brems only} & Side\\
		\textit{FEP clean} & (1620\,$\pm$\,5)\,keV & R$_{90}>$2\,mm, \textit{Comp} \& \textit{Brems only} & Top\\
		\midrule 		
	\end{tabular}
	\caption{\label{tab:libraries} MSE and SSE libraries used to evaluate the recognition efficiencies of ANNs. The energy range of the events contained in the libraries are given in the second column. The third column describes the selection criteria for the individual libraries, while information on the location of the simulated source, influencing the event location distribution, is listed in the fourth column. For details on the notation, see the text. } 
\end{table*}
\begin{figure*}
	\centering \subfigure[]{ \epsfig{file=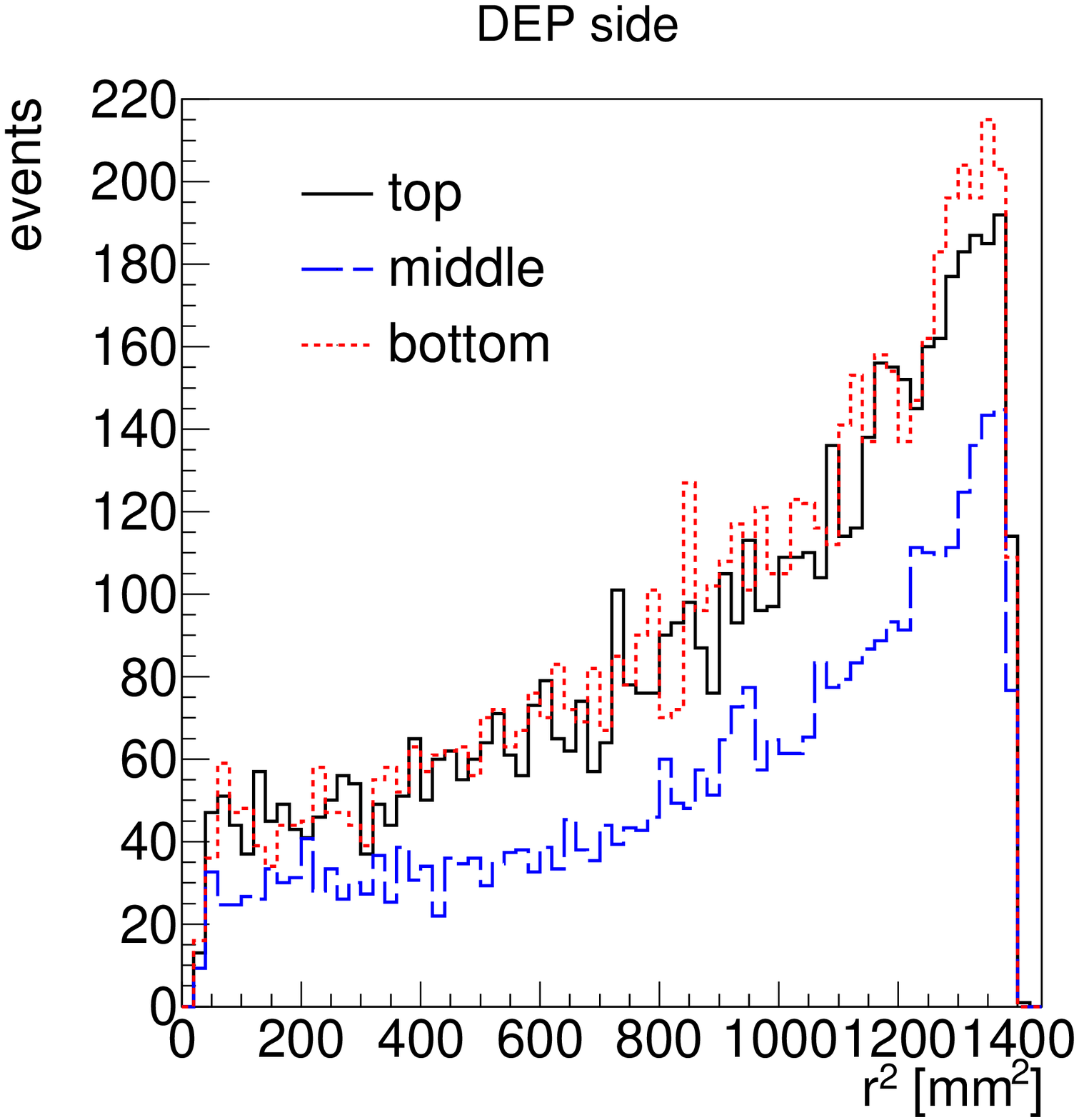,width=0.45 
	\textwidth}\label{fig:psa:event_distr_a} } \subfigure[]{ \epsfig{file=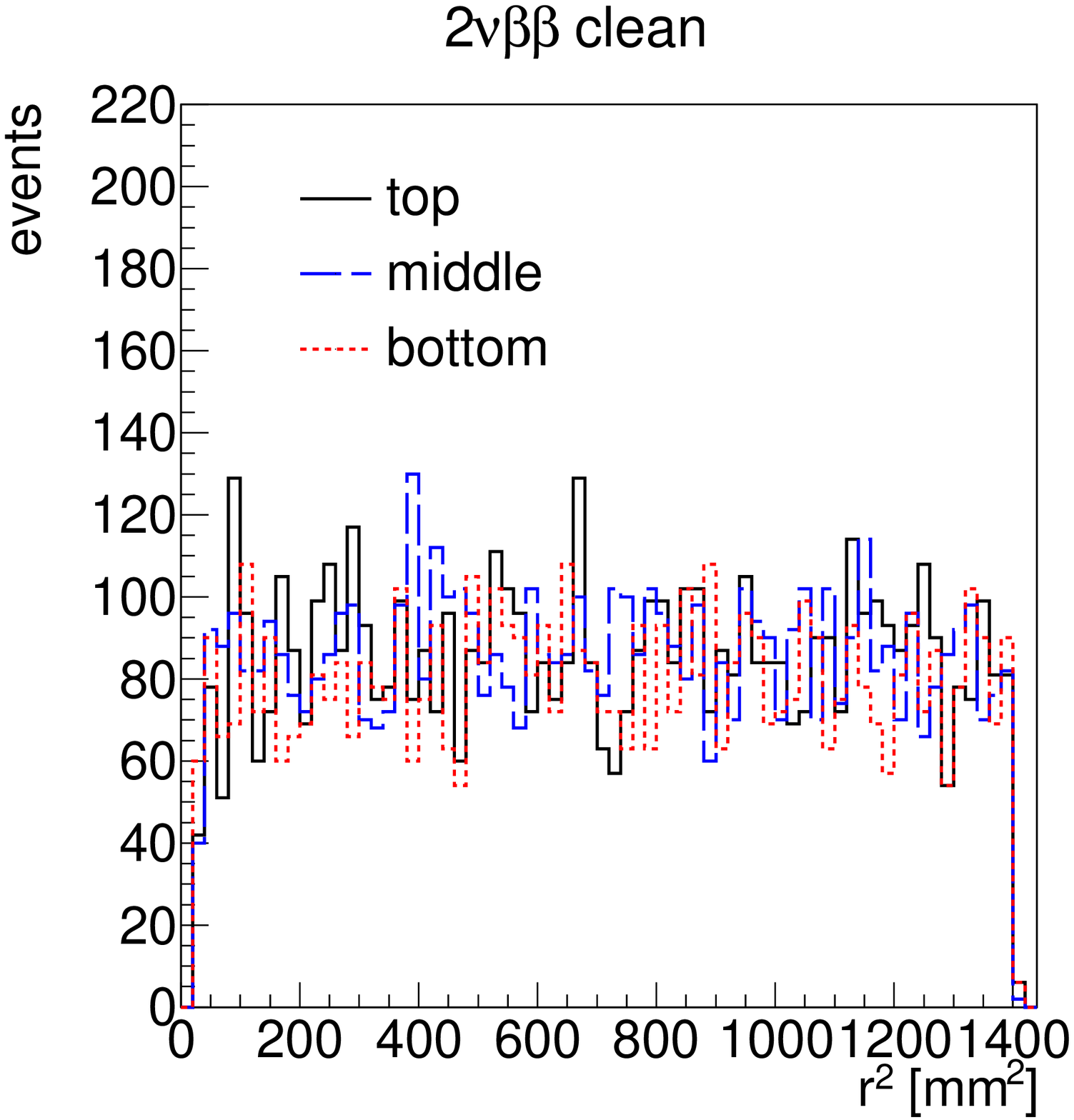,width=0.45 
	\textwidth}\label{fig:psa:event_distr_b} } \subfigure[]{ \epsfig{file=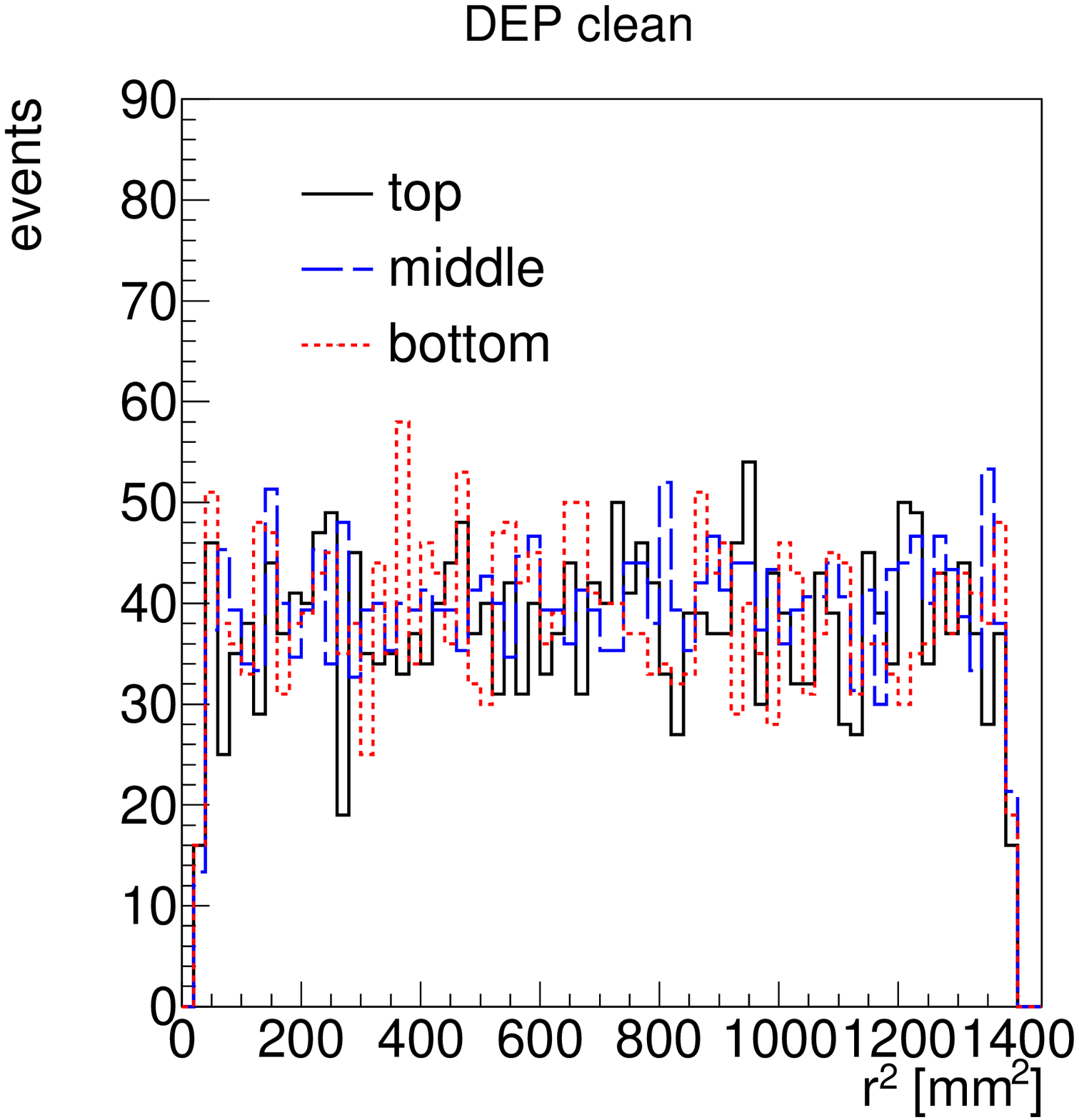,width=0.45 
	\textwidth}\label{fig:psa:event_distr_c} } \subfigure[]{ \epsfig{file=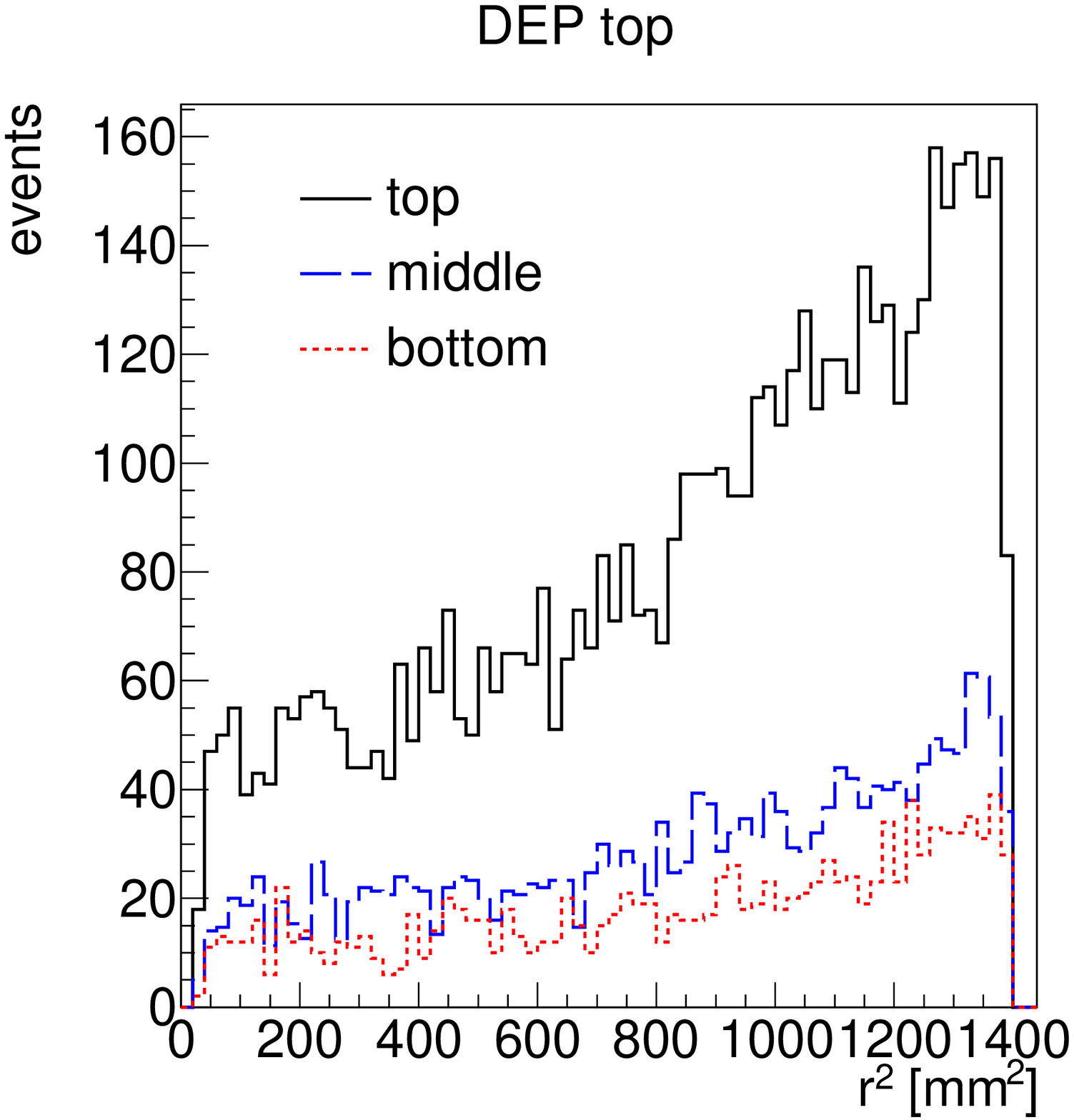,width=0.45 
	\textwidth}\label{fig:psa:event_distr_d} } \caption{Barycenter distributions of energy deposits in (a) \textit{DEP side}, (b) \textit{2$\nu\beta\beta$ clean}, (c) \textit{DEP clean} and (d) \textit{DEP top} libraries. The distributions, which are projections on the $xy$-plane, are plotted against the distance from the core squared, $r^2$, in order to ``normalize'' the distributions per unit area. Top, middle and bottom refer to events contained in the upper, middle and lower third of the detector, respectively. }\label{fig:psa:event_distr} 
\end{figure*}

The radial distributions of the energy barycenters, defined as the energy-weighted mean radial position of the energy deposit, of individual events for SSE libraries containing no MSEs are shown in Fig.~\ref{fig:psa:event_distr}. Top, middle and bottom refer to events contained in the upper, middle and lower third of the detector, respectively. These three volumes are equal. The barycenter of an individual event corresponds approximately to the position of the interaction/decay. For the \textit{DEP clean} library, where \textit{clean} is used here and below to identify libraries with no Compton or Bremsstrahlung interactions, it is flat as a function of $r$ and equivalent to the distribution of the \textit{2$\nu\beta\beta$ real} library. \textit{Real} is used to indicate libraries which contain all processes, i.e.\ including Compton scattering and Bremsstrahlung. The \textit{DEP side} and \textit{DEP top} libraries have inhomogeneous event-location distributions, events being located with a higher probability at high $r$ close to the bottom and top of the detector since for these parts of the detector it is more likely for the two back to back 511\,keV photons to escape the detector. \textit{Side} and \textit{top} indicate the location of the source with respect to the detector. 

\section{ANN training and efficiencies}\label{sec:training} The libraries listed in Table~\ref{tab:libraries} were used to create five different ANN training and evaluation sets each. They are listed in Table~\ref{tab:training_sets}, showing the combinations of SSE and MSE libraries. 
\begin{table*}
	\centering 
	\begin{tabular}
		{lcc} & SSE library & MSE library\\
		\midrule {\it set I - inhom DEP\/} & {\it DEP side\/} & {\it FEP side\/}\\
		{\it set II - real 2$\nu\beta\beta$\/} & {\it 2$\nu\beta\beta$ real\/} & {\it FEP top\/}\\
		{\it set III - hom DEP\/} & {\it DEP clean\/} & {\it FEP top\/}\\
		{\it set IV - top DEP\/} & {\it DEP top\/} & {\it FEP top\/}\\
		{\it set V - clean 0$\nu\beta\beta$\/} & {\it 0$\nu\beta\beta$ clean\/} & {\it FEP top\/}\\
		\midrule 
	\end{tabular}
	\caption{\label{tab:training_sets} Sets of libraries used for ANN training and efficiency evaluations. \textit{Hom} and \textit{inhom} are used to indicate sets where a SSE library with homogeneous and inhomogeneous event location distribution inside the detector was used. Note that individual libraries with independent pulses were used for training and evaluation of the ANNs. } 
\end{table*}

The ANNs used in this analysis were built using the TMultiLayerPerceptron (TMLP) within the ROOT framework~\cite{ROOT}. Only the part of the pulse containing the relevant information on the event topology is used by the ANNs. The pulses in the considered detector are maximally around 500\,ns long. In total, 40 time steps, corresponding to 530\,ns, were used. The center of the resulting trace was chosen to be the point where the pulse reaches 50\,\% of its amplitude. The amplitude of each pulse was normalized to unity.

The ANNs are composed of 40 input neurons, one hidden layer with the same number of neurons and an output layer with only one neuron. The ANNs were trained using the Broyden-Fletcher-Goldfarb-Shanno learning method~\cite{BFGS_method,BFGS_method_2,BFGS_method_3,BFGS_method_4}. Background-like MSEs were assigned an ANN output, $NN$, of 0 and signal-like SSEs were assigned an $NN$ of 1. Libraries of the same size for MSEs and SSEs were used. For a trained network, $NN$ should be close to 1 for SSEs and close to 0 for MSEs. 

For each individual ANN events are classified as SSE if $NN>\overline{NN}$, where $\overline{NN}$ is a parameter that has to be optimized. The rejection probability $R(\overline{NN})$ represents the fraction of events from an MSE dominated library, FEP in this case, rejected by the cut $NN \leq \overline{NN}$. The survival probability $E(\overline{NN})$ represents the fraction of events from a SSE dominated library (DEP, $2\nu\beta\beta$ or $0\nu\beta\beta$) kept with $NN>\overline{NN}$. The cut value $\overline{NN}_{\max}$ is chosen for each individual ANN to maximize the quantity $\varepsilon = \sqrt{R(\overline{NN})\cdot E(\overline{NN})}$ of the corresponding evaluation set used. This ensures that the highest E and R are obtained at the same time. 
\begin{figure*}
	\centering \epsfig{file=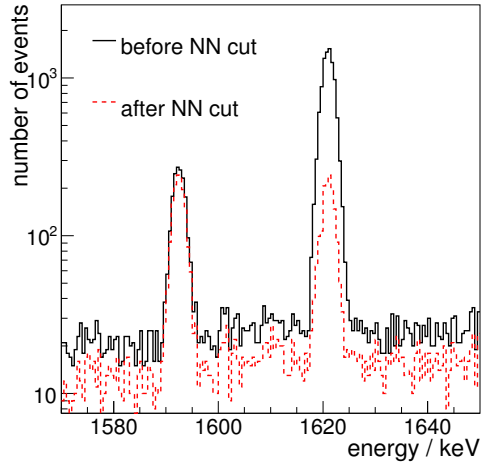,width=0.55 
	\textwidth} \caption{Simulated spectra of events contained in the libraries {\it DEP real\/} and {\it FEP all\/} (see Table~\ref{tab:libraries}) in the energy region around the 1593\,keV DEP and the 1620\,keV FEP before (solid line) and after (dashed line) MSE rejection using the ANN trained with {\it set I}.\label{fig:sample_spectrum} } 
\end{figure*}

The solid histogram in Fig.~\ref{fig:sample_spectrum} shows the simulated energy spectrum for events contained in the {\it DEP real\/} and {\it FEP all\/} libraries (Table~\ref{tab:libraries}). The FEP is significantly reduced while the DEP remains almost untouched.

The survival probability $E(\overline{NN})$ for 0$\nu\beta\beta$ and DEP events is given by the ratio of the peak areas after and before the ANN rejection. The areas are determined by fitting a Gaussian plus constant background to the spectra.

\begin{figure*}
	\centering \subfigure[]{ \epsfig{file=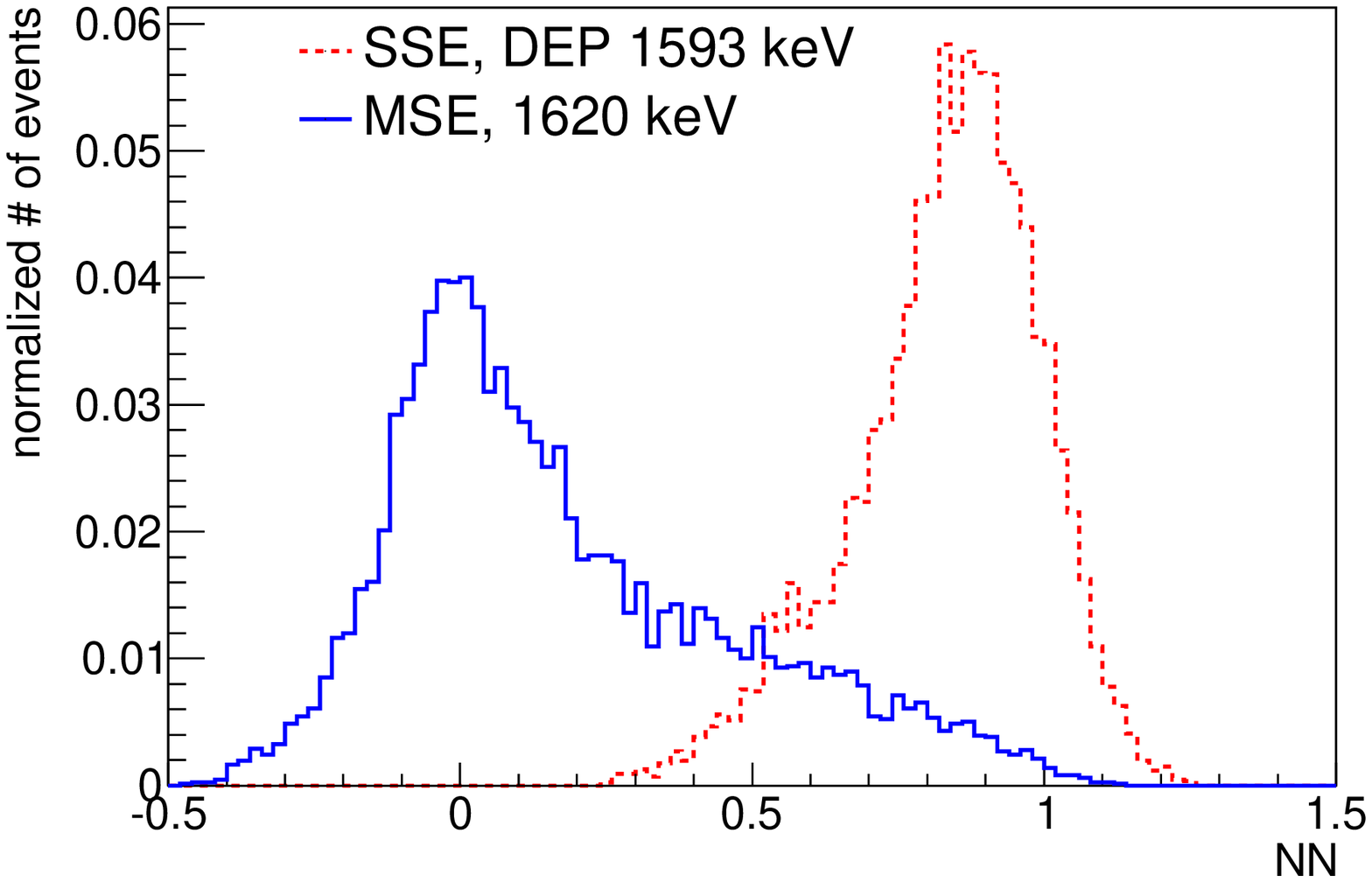,width=0.45 
	\textwidth, trim = 2mm 0mm 15mm 10mm, clip}\label{fig:psa:training_a} } \subfigure[]{ \epsfig{file=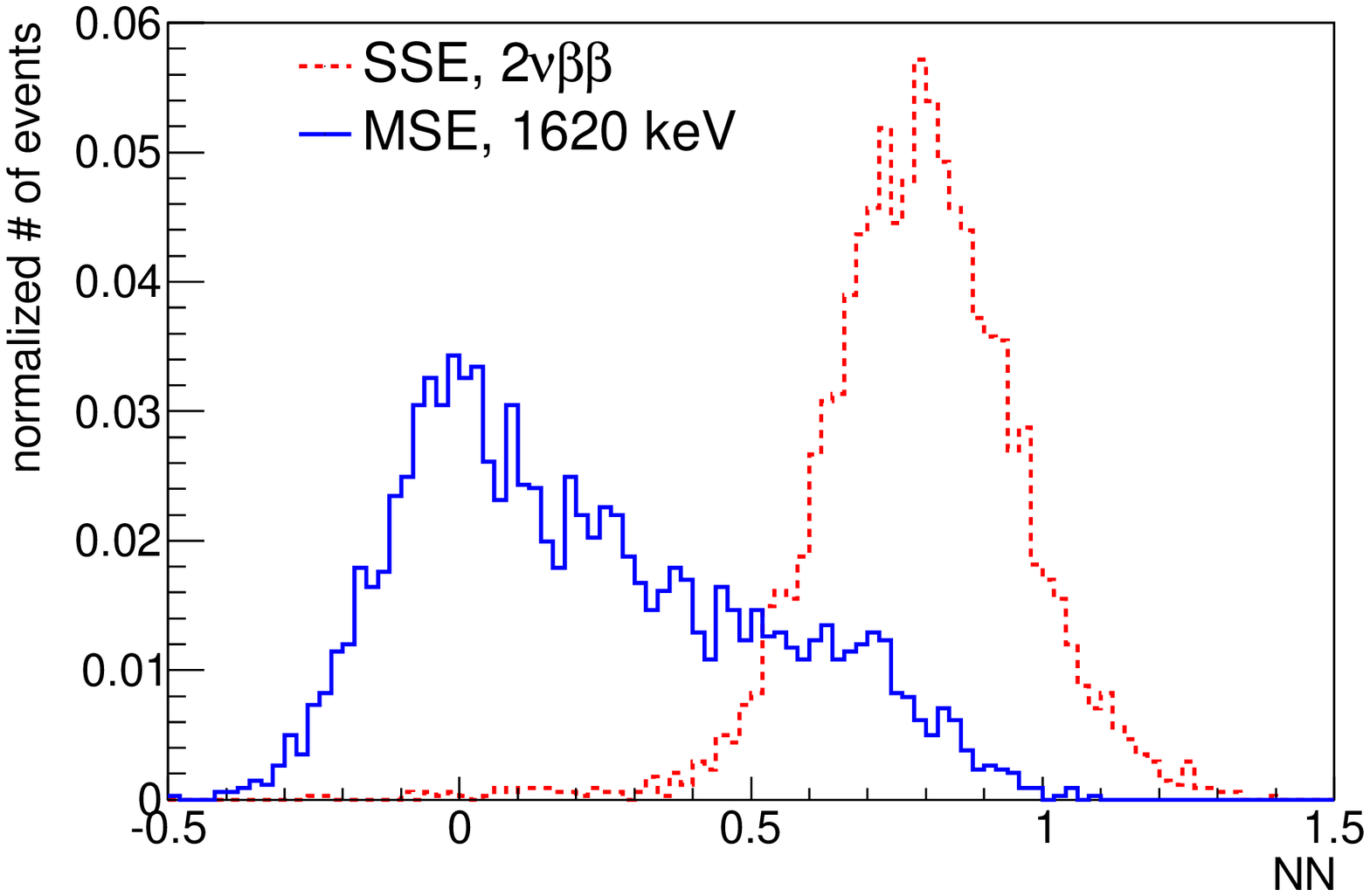,width=0.45 
	\textwidth, trim = 2mm 0mm 15mm 10mm, clip}\label{fig:psa:training_b} } \subfigure[]{ \epsfig{file=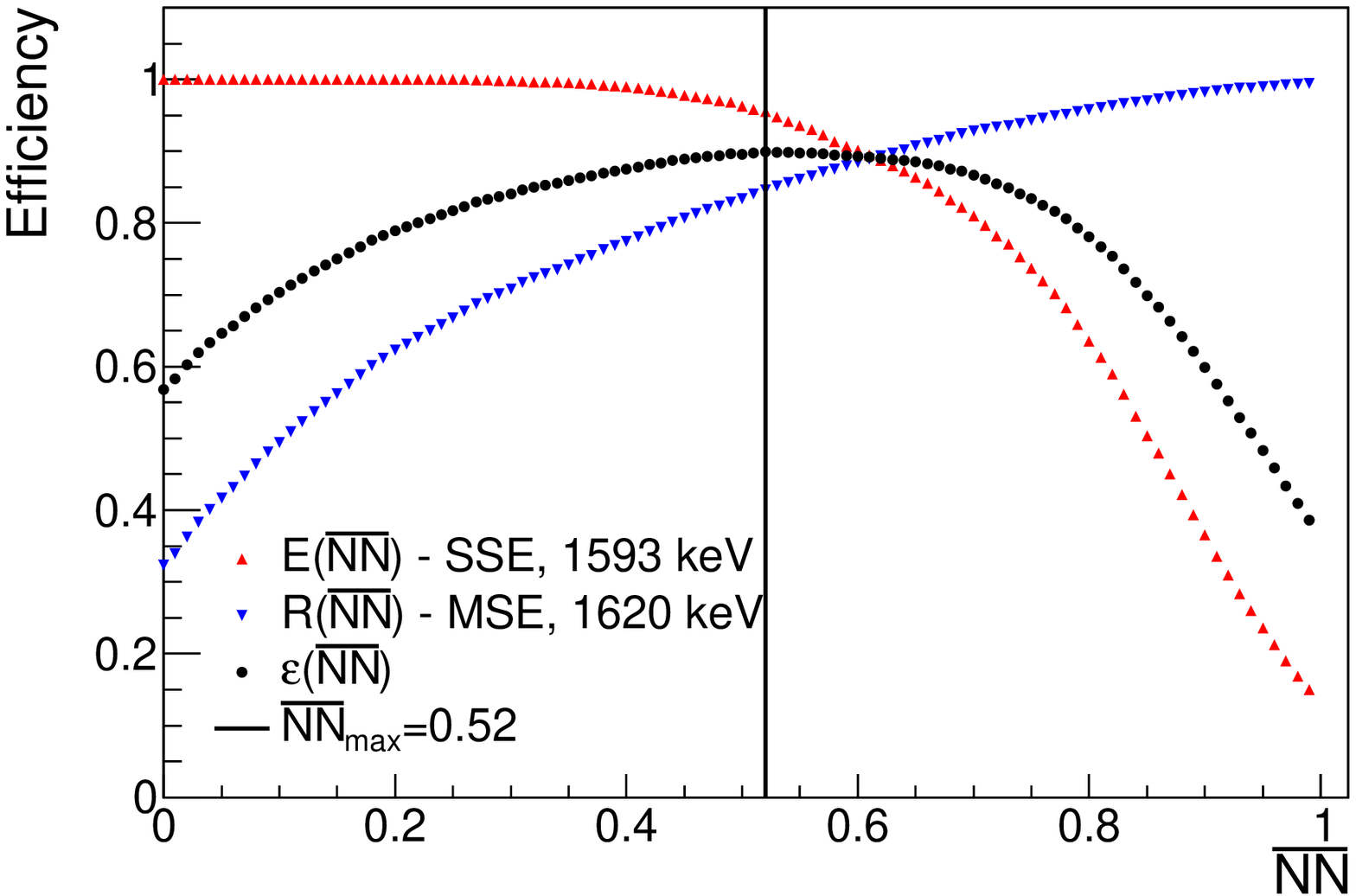,width=0.45 
	\textwidth, trim = 2mm 0mm 15mm 10mm, clip}\label{fig:psa:training_c} } \subfigure[]{ \epsfig{file=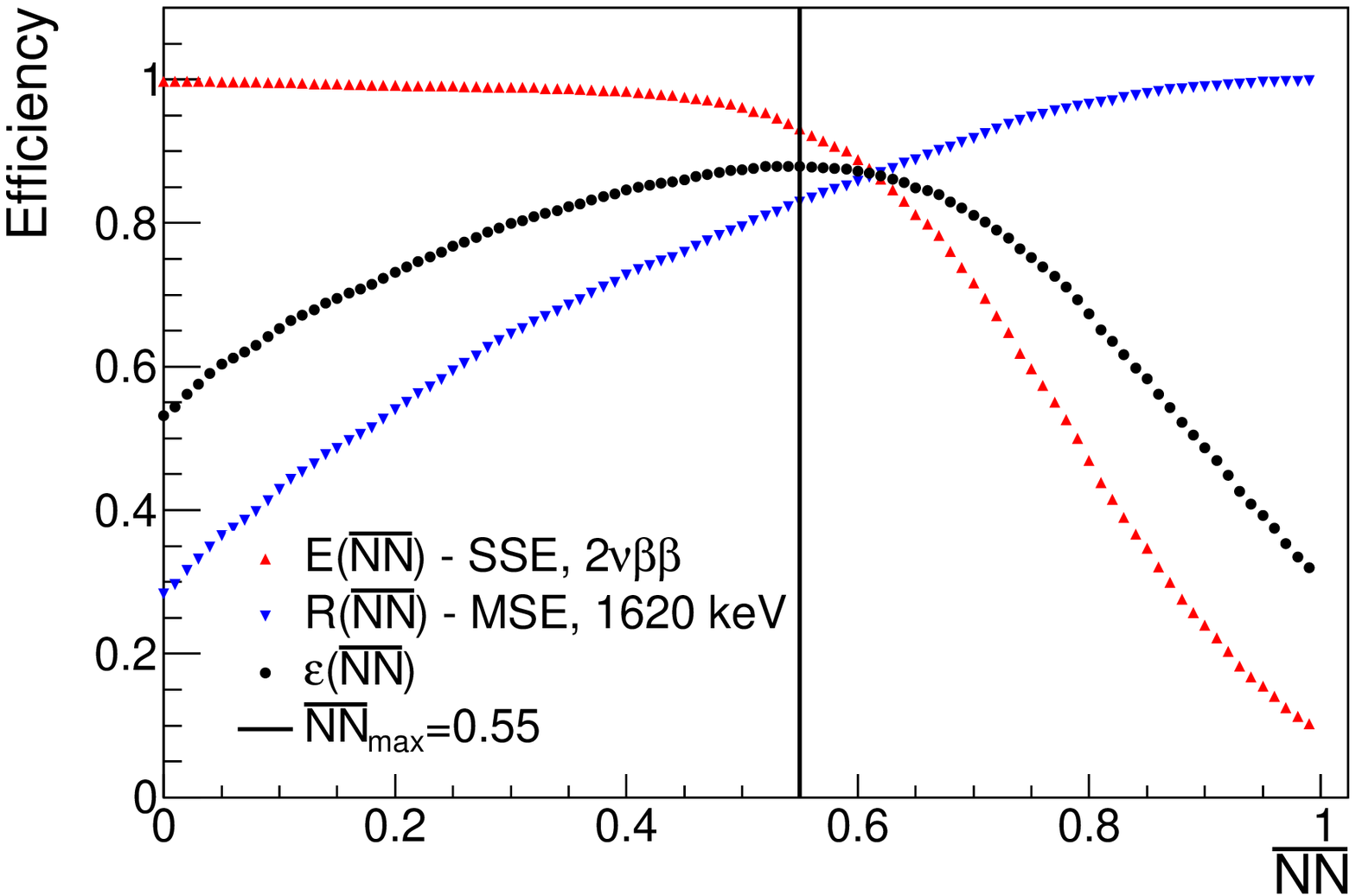,width=0.45
	\textwidth, trim = 2mm 0mm 15mm 10mm, clip}\label{fig:psa:training_d} } \caption{NN output distributions for SSEs and MSEs in \subref{fig:psa:training_a} training \textit{set~I} and \subref{fig:psa:training_b} training \textit{set~II}. Distributions of $E(\overline{NN})$, $R(\overline{NN})$ and $\varepsilon(\overline{NN})$ derived from \subref{fig:psa:training_c} training {\it set I\/} and \subref{fig:psa:training_d} training {\it set II}. The vertical lines represent the cut $\overline{NN}_{\max}$ to obtain $\varepsilon_{\max}$.\label{fig:psa:rej-eff} } 
\end{figure*}

In Figures~\ref{fig:psa:training_a} and~\ref{fig:psa:training_b}, the $NN$ distribution for an ANN trained with MSEs and SSEs from training \textit{sets I} and \textit{II} are shown. A clear separation between the $NN$ distributions of the MSE and SSE libraries is visible. Figures~\ref{fig:psa:training_c} and~\ref{fig:psa:training_d} show $E(\overline{NN})$, $R(\overline{NN})$ and $\varepsilon(\overline{NN}$) for training {\it sets I\/} and {\it II}, respectively. The vertical line represents ${\overline{NN}_{\max}}$. The $E(\overline{NN}_{\max})$, $R(\overline{NN}_{\max})$, $\varepsilon(\overline{NN}_{\max})$ and $\eta(\overline{NN}_{\max})$ values are called $E$, $R$, $\varepsilon$ and $\eta$, respectively, in the following. These variables are summarized in Table~\ref{tab:variables}. For libraries with purely SSE or MSE events, for which $S_{SSE}=1$ and $B_{MSE}=1$, $E$ and $R$ coincide with $\eta$ and $\rho$, respectively (see Eq.~\ref{equ:1}).

Note that the ANNs with the optimized ${\overline{NN}_{\max}}$ as described here are later used for efficiency and uncertainty evaluations. 
\begin{table*}
	\centering 
	\begin{tabular}
		{clcc} \midrule Survival probability & & $E = E(\overline{NN}_{\max})$\\
		Rejection probability & & $R = R(\overline{NN}_{\max})$\\
		Signal recognition efficiency & & $\eta = \eta(\overline{NN}_{\max})$\\
		Background recognition efficiency & & $\rho = \rho(\overline{NN}_{\max})$\\
		Background reduction power & & $\varepsilon = \varepsilon(\overline{NN}_{\max})$\\
		\midrule 
	\end{tabular}
	\caption{\label{tab:variables} Summary of the variables used to evaluate the performance of the ANNs.} 
\end{table*}

Statistical uncertainties quoted in the following are derived from the statistical fluctuations expected due to the limited number of simulated events and events surviving the selection. 

\section{Influence of initial conditions, ANN topology and training libraries on recognition efficiencies}\label{sec:conditions}

\subsection{Initial conditions and topologies} The reproducibility of $\eta$ was investigated by training five ANNs with the same ANN topology. The same training samples were used but the initial weights of the individual synapses of the untrained ANN were different in each case. Also the order in which individual pulses from the training sets were chosen for the iterative training was different for each ANN\@. The fluctuations of $\eta$ between the different ANNs are of the order of 1\% of the value of $\eta$, the RMS of the distribution of $\eta$ is taken as its systematic uncertainty. This systematic uncertainty only describes within which precision efficiencies are reproducible. They are not to be confused with systematic uncertainties related to pulse shape simulation.

Two groups of five ANNs, each with a different number of neurons in the hidden layer, were trained using the same training set (\textit{set II}). The value of $\eta$ for the default ANN with 40 hidden neurons was $0.976^{+0.001}_{-0.002}$ (stat.) $\pm0.005$ (syst.), while an ANN with 40 input neurons and one hidden layer with 10 neurons had a recognition efficiency $\eta=0.962^{+0.001}_{-0.002}$ (stat.) $\pm0.010$ (syst.). This is not significantly worse. Five ANNs with three hidden layers with 40 neurons each were trained and have $\eta=0.979^{+0.001}_{-0.001}$ (stat.) $\pm0.008$ (syst.). This is not a significant improvement with respect to the default ANN\@. The corresponding $\varepsilon$ values are ($0.917\pm0.013$), ($0.905\pm0.037$) and ($0.933\pm0.020$) for the default network, the network with one hidden layer of 10 neurons and the network with three hidden layers, respectively. In summary, the variation of $\eta$ due to the choice of the topology of the network is $^{+0.003}_{-0.014}$. In the following, the default network with one hidden layer with 40 neurons is used and the variation due to the topology is not considered in the following uncertainties.

\subsection{Recognition efficiencies as a function of training sample} ANNs were trained with the training sets listed in Table~\ref{tab:training_sets}. The trained ANNs were applied to the libraries $0\nu\beta\beta\ clean$ and \textit{FEP clean}, containing purely SSEs and MSEs. In this case, $S_{SSE}=1$ and $B_{MSE}=1$, respectively. Hence, $E_{\max}(S_{SSE}=1)$ = $\eta$ and $R_{\max}(B_{MSE}=1)$ = $\rho$ for a clean library (see Eq.~\ref{equ:1}). The resulting $\eta$, $\rho$ and $\varepsilon$ values are given in Table~\ref{tab:efficiencies}. 
\begin{table*}
	\centering 
	\begin{tabular}
		{lccc} \midrule Training set& $\eta$ ({\it $0\nu\beta\beta$ clean\/}) & $\rho$ ({\it FEP clean\/}) & $\varepsilon$ \\
		\midrule \textit{set I -- inhom DEP}& 0.915$\pm$0.017 & 0.893$\pm$0.014& 0.904\\
		\textit{set II -- real 2$\nu\beta\beta$} & 0.976$\pm$0.005& 0.862$\pm$0.008& 0.917\\
		\textit{set III -- hom DEP} & 0.964$\pm$0.009 & 0.887$\pm$0.006 & 0.924\\
		\textit{set IV -- top DEP} & 0.921$\pm$0.012 &0.888$\pm$0.006 & 0.904 \\
		\textit{set V -- clean 0$\nu\beta\beta$}&0.958$\pm$0.008 & 0.888$\pm$0.008& 0.922\\
		\midrule 
	\end{tabular}
	\caption{\label{tab:efficiencies} Signal recognition efficiency $\eta$, background recognition efficiency $\rho$, and $\varepsilon$ for ANNs trained with library sets having different SSE samples. Only the systematic uncertainties are quoted. The statistical uncertainties are for all numbers less than $\pm0.003$.} 
\end{table*}

The highest values for $\varepsilon$ were obtained with ANNs trained with SSE samples with homogeneous event-location distributions. The $\varepsilon$ values for ANNs trained with inhomogeneous samples are by approximately 0.02 lower. The variation on $\eta$ is up to 0.06 and hence more pronounced than on $\rho$ ($\approx$0.03). The variations due to SSE libraries with different event-location distributions used for the ANN training are significantly bigger than the fluctuations due to changes of the ANN initial conditions.

\section{$\boldsymbol{0\nu\beta\beta}$ detection efficiencies}\label{sec:uncertainties} 
\subsection{Survival probabilities for realistic $\boldsymbol{0\nu\beta\beta}$ and DEP samples} The survival probabilities $E$ obtained with the trained and ${\overline{NN}_{\max}}$ optimized ANNs were evaluated on the SSE libraries \textit{$2\nu\beta\beta$\ real}, \textit{$0\nu\beta\beta$\ real} and \textit{DEP real} according to the method explained in Sec.~\ref{sec:training}. The results are listed in Table~\ref{tab:Survival_Prob} for training $sets\ I$, $II$ and $V$. 
\begin{table*}\footnotesize
	\centering 
	\begin{tabular}
		{cccc} \midrule \diaghead(-4,1){clean $0\nu\beta'beta$ set bla bla} {\scriptsize{Evaluation library}}{\scriptsize{Training set}} & \shortstack{{\it set I\/} \\
		{\it inhom DEP\/}} & \shortstack{{\it set II\/} \\
		{\it real 2$\nu\beta\beta$\/}} & \shortstack{{\it set V\/} \\
		{\it clean 0$\nu\beta\beta$\/}} \\
		\midrule \textit{0$\nu\beta\beta$ real} & 0.867$^{+0.002}_{-0.003}\pm$0.018 & 0.937$^{+0.003}_{-0.004}\pm$0.005 &0.916$^{+0.003}_{-0.004}\pm$0.009\\
		\parbox[c]{2cm}{\textit{\centerline{2$\nu\beta\beta$ real}} \\
		\textit{\centerline{(\scriptsize{\text{1000\,keV\,$<$\,E\,$<$\,1450\,keV}})}}} & 0.885$^{+0.005}_{-0.007}\pm$0.017 & 0.944$^{+0.004}_{-0.005}\pm$0.005 & 0.915$^{+0.004}_{-0.006}\pm$0.008\\
		\textit{DEP real} & 0.898$^{+0.011}_{-0.016}\pm$0.012 & 0.936$^{+0.003}_{-0.001}\pm$0.003 &0.914$^{+0.011}_{-0.016}\pm$0.007\\
		\midrule $\Delta E(2\nu\beta\beta-0\nu\beta\beta)$ & (2.1$^{+0.6}_{-0.9}\pm$0.6)\% & (0.8$^{+0.5}_{-0.7}\pm$0.1)\%& (0.3$^{+0.5}_{-0.8}\pm$0.4)\%\\
		$\Delta E(\rm DEP-0\nu\beta\beta)$ & (3.5$^{+1.3}_{-1.9}\pm$1.1)\% & (-0.1$^{+1.0}_{-1.6}\pm$0.7)\%& (-0.2$^{+1.2}_{-1.8}\pm$0.6)\%\\
		\midrule 
	\end{tabular}
	\caption{\label{tab:Survival_Prob} The survival probability $E$ for the $DEP\ real$, {\it $2\nu\beta\beta$ real\/} and {\it $0\nu\beta\beta$ real\/} libraries from ANNs trained with \textit{sets I}, \textit{II} and \textit{V}. The differences of $E$ evaluated with the {\it DEP real\/} and {\it $2\nu\beta\beta$ real\/} samples to $E$ evaluated using the {\it $0\nu\beta\beta$ real\/} sample are also listed. Statistical and systematic uncertainties are quoted separately.} 
\end{table*}

For 0$\nu\beta\beta$ events in the energy interval $(2039\pm5)$~\,keV, $E_{\max}$ values of (0.937$\pm$0.006) and (0.867$\pm$0.018) were obtained with the ANN trained with \textit{set~II} and \textit{set~I}, respectively, where the statistical and systematic uncertainties were added in quadrature. For the ANN trained on \textit{set I}, $E$ is lower than for \textit{sets II} and \textit{V}, as expected from the lower $\eta$ for this training set (see Table~\ref{tab:efficiencies}). The realistic signal-like libraries also contain a significant amount of MSEs. This explains why the obtained $E$ values for 0$\nu\beta\beta$ are significantly different from the $\eta$ listed in Table~\ref{tab:efficiencies}. As the amount of wrong type of events in event libraries depends on geometry and energy this also implies that $E$ by itself is not a precise quantity to compare PSD methods even if $R$ is also considered. 

\subsection{Inhomogeneity of signal recognition efficiency} 
\begin{figure*}
	\centering \subfigure[]{ 
	\includegraphics[width=0.45 
	\textwidth]{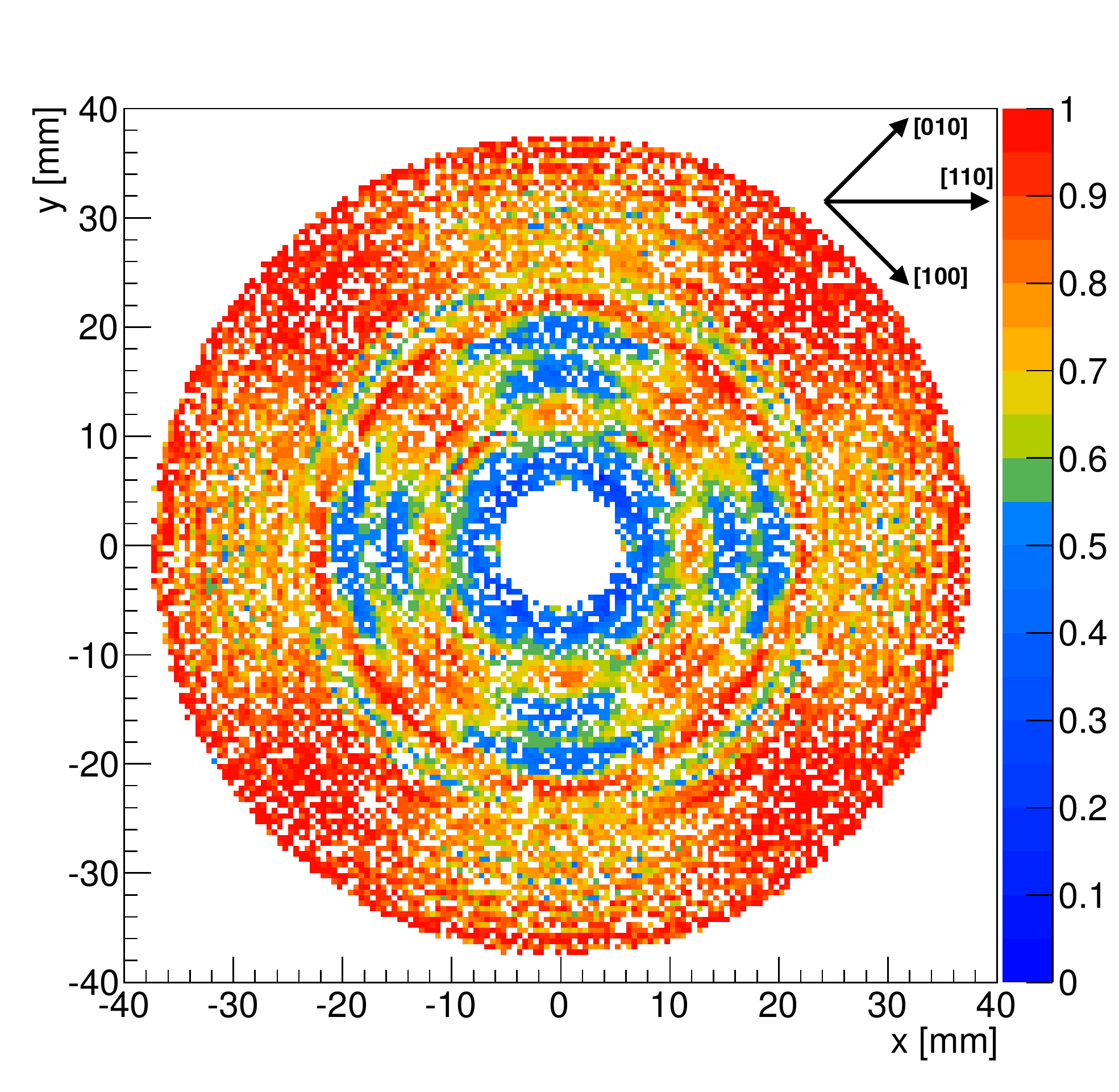}\label{fig:psa:XYaver_side} } \subfigure[]{ 
	\includegraphics[width=0.45 
	\textwidth]{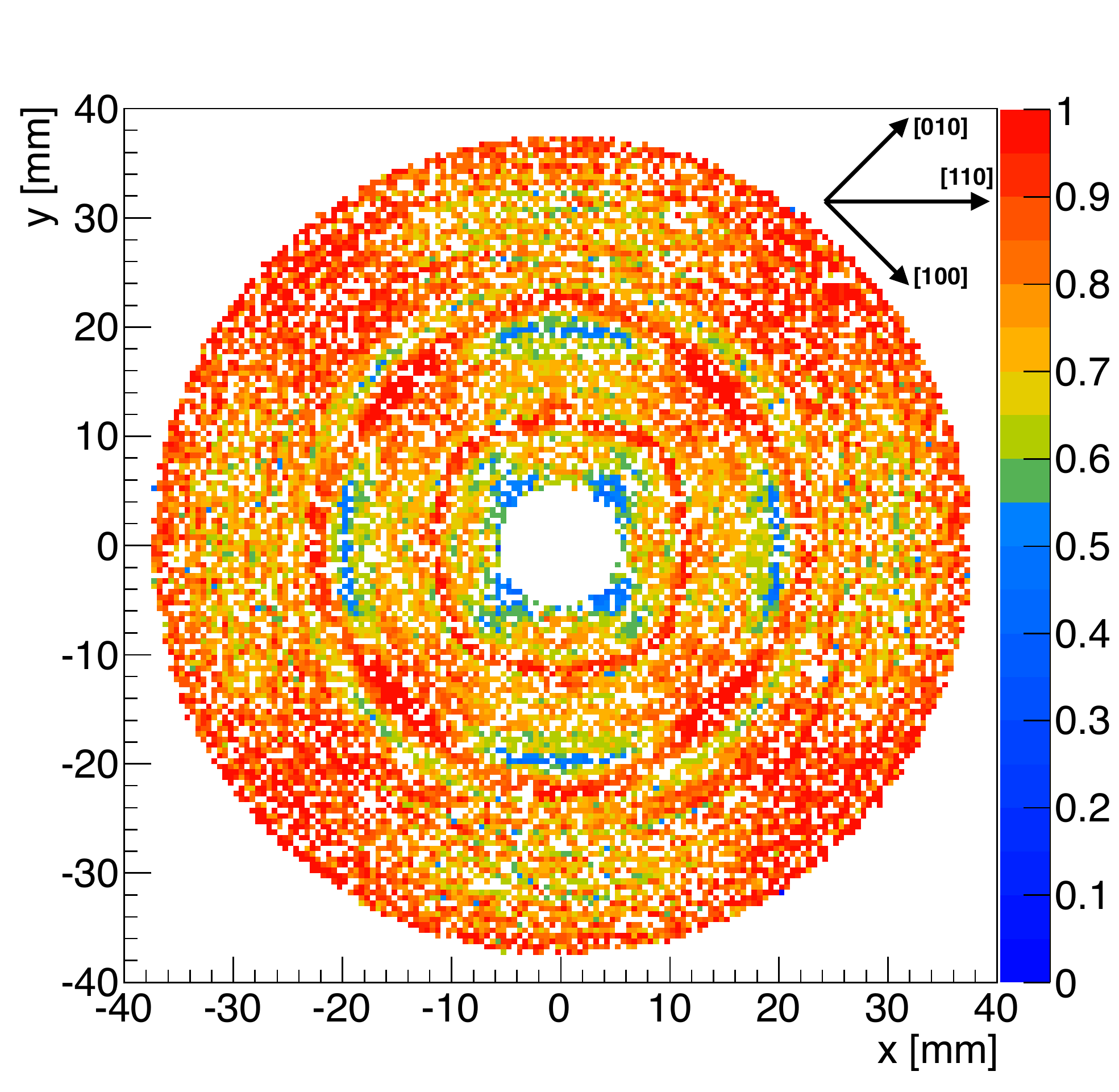}\label{fig:psa:XYaver_0nbbt} } \caption{Average NN output value (color palette) for events from the {\it $0\nu\beta\beta$ clean\/} library at different positions within the detector for ANNs trained with training \subref{fig:psa:XYaver_side} \textit{set~I} and \subref{fig:psa:XYaver_0nbbt} \textit{set V}. The orientation of the crystal axes are shown on the figure.}\label{fig:psa:XYaver}\vspace{2.5mm} 
\end{figure*}

The position distribution of the rejected events inside the detector, i.e.\ the position dependence of the signal recognition efficiency was studied. In Fig.~\ref{fig:psa:XYaver}, the location dependence of the mean value of the $NN$ output inside the detector is depicted for the SSEs from the \textit{0$\nu\beta\beta$ clean} library.

Regions where the average $NN$ output is lower than $\overline{NN}_{\max}=0.55$ are seen as blue areas. In these regions, SSEs are systematically rejected. The fraction of the volume where the SSEs of the {\it 0$\nu\beta\beta$ clean\/} library are more likely to be rejected than to be accepted as SSE is (8.0$\pm$1.7)\% for an ANN trained with the SSE sample with inhomogeneous event-location distribution \textit{set I}. For the ANNs trained with \textit{sets II} and \textit{V}, the affected volume is reduced to (2.2$\pm$0.5)\% and (3.7$\pm$0.8)\%, respectively~\cite{aleks_thesis}. Using an ANN training set with similar event-location distribution as for the evaluation set decreases the effect of the systematic volume cut, however, it does not completely remove it. 

The symmetry in the patterns observed in Fig.~\ref{fig:psa:XYaver} seems to be connected to the crystallographic symmetry of the detector. The axis dependence of the effect might be due to the dependence of the electron to hole mobility ratio on the position of the charge carriers with respect to the crystal axes (see Fig. 2 in~\cite{PSS}). Affected zones appear close to the inner detector surface and in the middle of the bulk around $r\approx18$\,mm. The mechanism of pattern formation is, however, not understood. 

\subsection{Consequences for $\boldsymbol{0\nu\beta\beta}$ analyses} The different event-location distributions for DEP samples from calibration and 0$\nu\beta\beta$ signal events (see Fig.~\ref{fig:psa:event_distr}) was identified as the major source of systematic uncertainty for the approach of ANNs trained with DEP sets. For 2$\nu\beta\beta$ training samples, the different energy distribution leads to a different signal-to-noise ratio. The $\eta$ values obtained for different SSE evaluation libraries with ANNs trained with different training sets are listed in Table~\ref{tab:efficiency_E}.

The signal recognition efficiencies $\eta$ of the different ANNs are within uncertainties the same for the different evaluation libraries with homogenous event-location distribution. This demonstrates that the normalization of the input to the ANN makes the influence of the lower energy of events, down to 1\,MeV, insignificant. However, when $\eta$ is derived using the \textit{DEP side} set with realistic event-location distribution it is systematically overestimated. There is a $\Delta\eta[set\ I](DEP\ side - 0\nu\beta\beta)=(4.2^{+0.6}_{-0.9}\pm0.8)\%$ effect for the ANN trained with an independent \textit{DEP side} SSE training set. For ANNs trained with homogeneous samples the effect is reduced but nevertheless significant with $\Delta\eta[set\ II](DEP\ side - 0\nu\beta\beta)=(1.1^{+0.3}_{-0.4}\pm0.7)\%$ and $\Delta\eta[set\ V](DEP\ side - 0\nu\beta\beta)=(1.4^{+0.4}_{-0.7}\pm0.9)\%$. Note that the efficiencies obtained when training the network with the $2\nu\beta\beta$ set are higher than the ones obtained using the $0\nu\beta\beta$ set, the reason for which is unclear.

Comparing the resulting $\eta$ with $E$ quoted in Table~\ref{tab:Survival_Prob} shows that the additional admixture of MSEs to the evaluation libraries slightly reduces $E$ with respect to $\eta$. 
\begin{table*}\footnotesize
	\centering 
	\begin{tabular}
		{cccc} \midrule \diaghead(-4,1){clean $0\nu\beta'beta$ set bla bla} {\scriptsize{Evaluation library}}{\scriptsize{Training set}} & \shortstack{{\it set I\/} \\
		{\it inhom DEP\/}} & \shortstack{{\it set II\/} \\
		{\it real 2$\nu\beta\beta$\/}} & \shortstack{{\it set V\/} \\
		{\it clean 0$\nu\beta\beta$\/}} \\
		\midrule \textit{$0\nu\beta\beta$ clean }& 0.915$^{+0.003}_{-0.004}\pm$0.017 & 0.976$^{+0.001}_{-0.001}\pm$0.005& 0.958$^{+0.001}_{-0.002}\pm$0.008\\
		\textit{$2\nu\beta\beta$ clean } & 0.911$^{+0.005}_{-0.007}\pm$0.018 & 0.970$^{+0.003}_{-0.004}\pm$0.007 & 0.956$^{+0.004}_{-0.006}\pm$0.008\\
		\textit{DEP clean} & 0.917$^{+0.004}_{-0.006}\pm$0.018 & 0.976$^{+0.002}_{-0.002}\pm$0.005 & 0.960$^{+0.002}_{-0.003}\pm$0.009\\
		\textit{DEP side} & 0.954$^{+0.004}_{-0.007}\pm$0.011 & 0.987$^{+0.003}_{-0.004}\pm$0.004 & 0.971$^{+0.004}_{-0.006}\pm$0.006\\
		\midrule $\Delta\eta(2\nu\beta\beta - 0\nu\beta\beta$) & (-0.4$^{+0.6}_{-0.9}\pm$0.1)\% & (-0.6$^{+0.3}_{-0.4}\pm$0.3)\%& (-0.2$^{+0.4}_{-0.7}\pm$0.2)\%\\
		$\Delta\eta(DEP\ clean - 0\nu\beta\beta$) & (0.1$^{+0.5}_{-0.8}\pm$0.2)\% & (0.0$\pm$0.2$\pm$0.03)\%& (0.2$^{+0.2}_{-0.4}\pm$0.1)\%\\
		$\Delta\eta(DEP\ side - 0\nu\beta\beta$) & (4.2$^{+0.6}_{-0.9}\pm$0.8)\% & (1.1$^{+0.3}_{-0.4}\pm$0.7)\%& (1.4$^{+0.4}_{-0.7}\pm$0.4)\%\\
		\midrule 
	\end{tabular}
	\caption{\label{tab:efficiency_E} Values of $\eta$ for different libraries for ANNs trained with different sets (see Table~\ref{tab:training_sets}). The difference in efficiency between evaluation sets is also listed.} 
\end{table*}

\section{Summary and conclusions}\label{sec:conclusions} Systematic effects on the determination of the signal recognition efficiency of pulse-shape-analysis using ANNs were investigated using pulse shape simulation. The most important effect was found to be due to the event-location distribution of the evaluation libraries. In contrast, the energy distribution of events in the training library was found to be irrelevant within reasonable limits.

It was found that training with SSE libraries with homogeneous event location distributions lead to higher signal recognition efficiencies. 

The use of evaluation libraries with homogeneous event location distribution lead to reduced systematic uncertainties on the signal recognition efficiencies of the order of 1\%. On the contrary signal recognition efficiencies of ANNs determined from DEP libraries with inhomogeneous event location distributions were found to be up to 5\% too high, consistent with the systematic uncertainties derived in~\cite{gerda_psa}. Differences in the energy distribution of the events of the evaluation samples do not have a significant effect. The different event-location distributions resulting from different positions of the calibration sources may result in variations of the ANN signal recognition efficiency by up to 6\% and the background discrimination power by 2\%. 

The signal detection efficiency of an ANN depends on the location of the events inside a true-coaxial detector. The efficiency is above 90\% in most parts of the detector. However, SSEs in the inner regions and in the center of the bulk are systematically misidentified. About 2\% to 8\% of the volume is affected, depending on the homogeneity of the event-location distribution of the training set used. Using training sets with homogeneous SSE location distribution reduces the affected regions but does not eliminate them completely.

The true-coaxial detectors assumed for these studies have particularly simple field configurations. The effects on detectors with more complex field configurations will have to be studied very carefully.

Pulse-shape discrimination with artificial neural networks is a useful tool to identify multi-site events. It potentially increases the sensitivity of 0$\nu\beta\beta$ experiments like GERDA~\cite{gerda,gerda_0nbb}. The usage of $2\nu\beta\beta$ events for training and efficiency evaluation of the artificial neural networks is recommended.

\end{document}